\documentclass{aastex701}

\begin{document}

\title{Evolution of Flare Ribbon Bead-like Structures in a Solar Flare}

\author[0000-0001-9726-0738]{Ryan J. French}
\affiliation{Laboratory for Atmospheric and Space Physics, University of Colorado Boulder, Boulder, CO 80303, USA}
\email{ryan.french@lasp.colorado.edu}

\author[0000-0001-8975-7605]{Maria D. Kazachenko}
\affiliation{Laboratory for Atmospheric and Space Physics, University of Colorado Boulder, Boulder, CO 80303, USA}
\affiliation{National Solar Observatory, 3665 Discovery Drive, 80303, Boulder, CO, USA}
\affiliation{Dept. of Astrophysical and Planetary Sciences, University of Colorado, Boulder, 2000 Colorado Ave, 80305, Boulder, CO, USA}
\email{Maria.Kazachenko@lasp.colorado.edu}

\author[0000-0003-4052-9462]{David Berghmans}
\affiliation{Solar-Terrestrial Centre of Excellence—SIDC, Royal Observatory of Belgium, 3 Avenue Circulaire, B-1180 Uccle, Belgium}
\email{david.berghmans@sidc.be}

\author[0000-0002-2914-2040]{Elke D’Huys}
\affiliation{Solar-Terrestrial Centre of Excellence—SIDC, Royal Observatory of Belgium, 3 Avenue Circulaire, B-1180 Uccle, Belgium}
\email{elke.dhuys@oma.be}

\author[0000-0002-1196-4046]{Marie Dominique}
\affiliation{Solar-Terrestrial Centre of Excellence—SIDC, Royal Observatory of Belgium, 3 Avenue Circulaire, B-1180 Uccle, Belgium}
\email{marie.dominique@oma.be}

\author[0000-0001-8504-2725]{Ritesh Patel}
\affiliation{Southwest Research Institute, 1301 Walnut Street, Suite 400, Boulder, CO 80302, USA}
\email{ritesh.patel@swri.org}

\author[0000-0002-9311-9021]{Dana-Camelia Talpeanu}
\affiliation{Solar-Terrestrial Centre of Excellence—SIDC, Royal Observatory of Belgium, 3 Avenue Circulaire, B-1180 Uccle, Belgium}
\email{dana.talpeanu@observatory.be}

\author[0000-0002-3229-1848]{Cole A. Tamburri}
\affiliation{National Solar Observatory, 3665 Discovery Drive, 80303, Boulder, CO, USA}
\affiliation{Dept. of Astrophysical and Planetary Sciences, University of Colorado, Boulder, 2000 Colorado Ave, 80305, Boulder, CO, USA}
\affiliation{Laboratory for Atmospheric and Space Physics, University of Colorado Boulder, Boulder, CO 80303, USA}
\email{cole.tamburri@colorado.edu}

\author[0000-0003-4065-0078]{Rahul Yadav}
\affiliation{Laboratory for Atmospheric and Space Physics, University of Colorado Boulder, Boulder, CO 80303, USA}
\affiliation{National Solar Observatory, 3665 Discovery Drive, 80303, Boulder, CO, USA}
\email{rahul.yadav@lasp.colorado.edu}

\begin{abstract}

We present fast cadence and high resolution observations of flare ribbons from the Solar Orbiter Extreme Ultraviolet Imager (EUI). Utilizing the short-exposure observations from the EUI High Resolution Imager in EUV (HRIEUV), we find small-scale blob/bead-like kernel structures propagating within a hook at the end of a flare ribbon, during the impulsive phase of a C9.9-class solar flare. These bead structures are dynamic, with well-resolved spatial separations as low as $\approx 420-840$ kilometers (3-6 pixels) -- below the observable limit of full-disk solar imagers. We analyze the evolution of the plane-of-sky apparent velocity and separation of the flare ribbon structures, finding evidence for multiple processes occurring simultaneously within the flare ribbon. These processes include -- quasi-periodic pulsation (QPP)-like brightenings, slow back-and-forth zig-zag motions along the ribbon, rapid apparent motions along the ribbon (600+ km/s), and stationary blob-like structures. Finally, we conduct Fast Fourier Transform  analysis and analyze the start times of exponential growth in the power spectrum at different spatial scales across the flare ribbon. Our analysis reveals that the ribbon beads form with a key spatial separation of 1.7-1.9 Mm, before developing into more complex structures at progressively larger and smaller spatial scales. This observation is consistent with predictions of the tearing mode instability.

\end{abstract}

\accepted{to ApJL, Nov 2025}


\section{Introduction} \label{sec:intro}

During solar flares, magnetic reconnection accelerates high-energy particles from the flare current sheet, down magnetic loops, towards the Sun's surface. As these high-energy particles reach the chromosphere, they deposit energy at the magnetic footpoints, heating local plasma to form flare ribbons \citep{Doschek1983, Cheng1983}. Flare ribbons, therefore, mark the magnetic connectivity between the chromosphere and flaring current sheet in the solar corona.
Bright ribbon features, known as kernels, are interpreted as the ribbon regions most immediately-connected to the energy release site \citep{Demoulin1993,Asai2002}.
Due to this magnetic connectivity, flare ribbon dynamics must in some way reflect the fundamental energy release processes within the flare current sheet \citep{Forbes2000}, a region challenging to probe directly. 
Observations of flare ribbons can therefore offer insights into the physics dictating flare onset and evolution. Flare ribbons are not smooth and laminar, but exhibit internal variations within them. Variations along flare ribbons are not caused by plasma motions along the flare ribbons themselves (at the scales observed by current EUV telescopes), but rather \textit{apparent motions} caused by changes in magnetic connectivity to the current sheet.
In recent years, there has been strong interest in investigating the origins of this \textit{sub-structure} or \textit{fine-structure} in flare ribbons. \textit{Sub-structure} (also spelled as \textit{substructure}) generally refers to variations across, along, or within the main ribbon structure. \textit{Fine-structure} typically refers to a sub-category of \textit{sub-structure}, referring to ribbon sub-structure near the observable limit of the highest spatial resolution telescopes. We note that these two terms are used somewhat interchangeably in some flare ribbon studies. 

The launch of the Interface Region Imaging Spectrograph \citep[IRIS,][]{DePontieu2014} led to a surge in flare ribbon studies from space-based observations. 
IRIS studies have found structures as small as $\approx150-300$ km in the flare ribbons \citep{Graham2015,Jing2016}, at the observable limit of the instrument. Ribbon structures of similar sizes have been found in ground-based H-alpha images from \citet{Sharykin2014,Thoen2025,Yaduv2025}, or even smaller by \citet{Tamburri2025}. The more recent of these studies refer to the bright ribbon kernel features as \textit{blobs}, and find them to be present for several minutes within the flare ribbon, without significant motion. Similar kernel structures have also been referred to as \textit{knots} \citep[][]{Sharykin2014,Sobotka2016}. As well as blob-like features, other studies have presented observations of hooks, swirls and fragmented fronts within the flare ribbons \citep[e.g.][]{Brannon2015,Li2015,Parker2017,Cannon2023,Polito2023}, spanning a wide range of velocities \citep[including super-Alfv\'enic apparent motions by][]{Lorincik2025}. In other recent works, vertical sub-structure along flare ribbons have been referred to as \textit{riblets} \citep{Singh2025}. 

Flare ribbon intensities, velocities and positions- have also been found to oscillate quasi-periodically in time \citep[e.g.][]{Brannon2015,Brosius2015,Naus2022,Corchado2024}. Although many current sheet processes have been proposed as the origins of ribbon sub-structure behaviors, many observable phenomena (including quasi-periodic pulsations, fragmented ribbon fronts and swirl/hook-like features) can be explained by the presence of the tearing mode instability in the coronal flare current sheet \citep{Biskamp1986}.  Simulation work also predicts the presence of the tearing mode instability in solar flares \citep[e.g.][]{Wyper2021,Dahlin2025}. Evidence for the tearing mode instability has also been detected in off-limb measurements of heated plasma around solar flare current sheets \citep[e.g.][]{Warren2018,Cheng2018,French2019,French2020}, but such observations are rare. Flare ribbon observations are far more numerous, and thus provide more frequent opportunities for probing current sheet processes.

%Historically flare ribbons have been classified to evolve in two stages: an initial parallel motion along the polarity inversion line (PIL), followed by a perpendicular expansion away from it \citep{Su2007,Qiu2009,Qiu2010,Cheng2012}. During the parallel stage, brightenings spread along the PIL either bidirectionally or unidirectionally (“zipper reconnection”; \citealt{Li2009,Qiu2017}) and often transition from high to low shear as the coronal guide field decreases \citep{Dahlin2021,Tamburri2024}. High-cadence IRIS observations revealed fine-scale, intermittent brightenings whose apparent slipping motions can reach thousands of km~s$^{-1}$—well above Alfvénic speeds—likely reflecting rapid sequential reconnection rather than plasma flows \citep{Lorincik2025}. In the subsequent perpendicular stage, ribbons move more slowly away from the PIL ($\lesssim$60~km~s$^{-1}$; \citealt{Hinterreiter2018}) or, in some cases, toward it due to reconnection between expanding flux-rope and overlying arcade fields \citep{Dudik2019,Zemanova2019,Aulanier2019}.

Magnetically speaking (although fundamentally different in their temperature, density and collisional timescales), flare ribbons are analogous to the auroral oval in the magnetosphere, where energy is deposited at the magnetic footpoints of the night-side magnetospheric current sheet. \textit{Auroral beads}, fine bead-like structures propagating along the aurora, as well as hook-and-swirl-like features, bear much resemblance to similar features in solar flares. These similarities provide interdisciplinary opportunities to share insights and methodologies between studies of these distinct plasma regimes. Inspired by methods used in auroral bead studies \citep{Rae2010,Kalmoni2015,Kalmoni2018}, \citet{French2021} found exponential growth in the power spectra at key spatial scales along the flare ribbon -- a key prediction of plasma instabilities. The exponential growth was found to begin at a key spatial scale, before the subsequent growth of progressively larger and smaller spatial scales. This is consistent with the simultaneous cascade and inverse-cascade of current sheet scales predicted by the tearing mode instability \citep{Tenerani2020}. 

In this work, we analyze observations of bead-like ribbon kernel structures by the Extreme Ultraviolet Imager \citep[EUI,][]{Rochus2020} onboard Solar Orbiter \citep{Muller2020}. We track intensity variations along the flare ribbon to reveal a multitude of bead behaviors, including rapid (600+ km/s) motions, quasi-periodic brightenings, slow zig-zag motions and stationary bright points. We compare these behaviors with expectations from the tearing mode instability. Finally, we employ the methodology from \citet{French2021} to track the start times of exponential growth at different spatial scales along the flare ribbon, to find evidence for a simultaneous cascade and inverse cascade of spatial scales in the flare current sheet.

\section{Observations} \label{sec:instruments}

Solar Orbiter observed a number of solar flares during its 2024 major flare campaigns \citep{Ryan2025}, including a C9.9-class flare on 2024 March 24th from AR 13615.  At the time of the flare, Solar Orbiter was at a distance of 0.38 au from the Sun. As a consequence of this shorter light travel distance/time, the instruments onboard Solar Orbiter observed the flare at a time 5 minutes and 7 seconds earlier than instruments at Earth. It is this Solar Orbiter reference time that we use in the figures and text contained within this work, not correcting for additional travel time to Earth.

\begin{figure*}%[h]
\centering
\includegraphics[width=18cm]{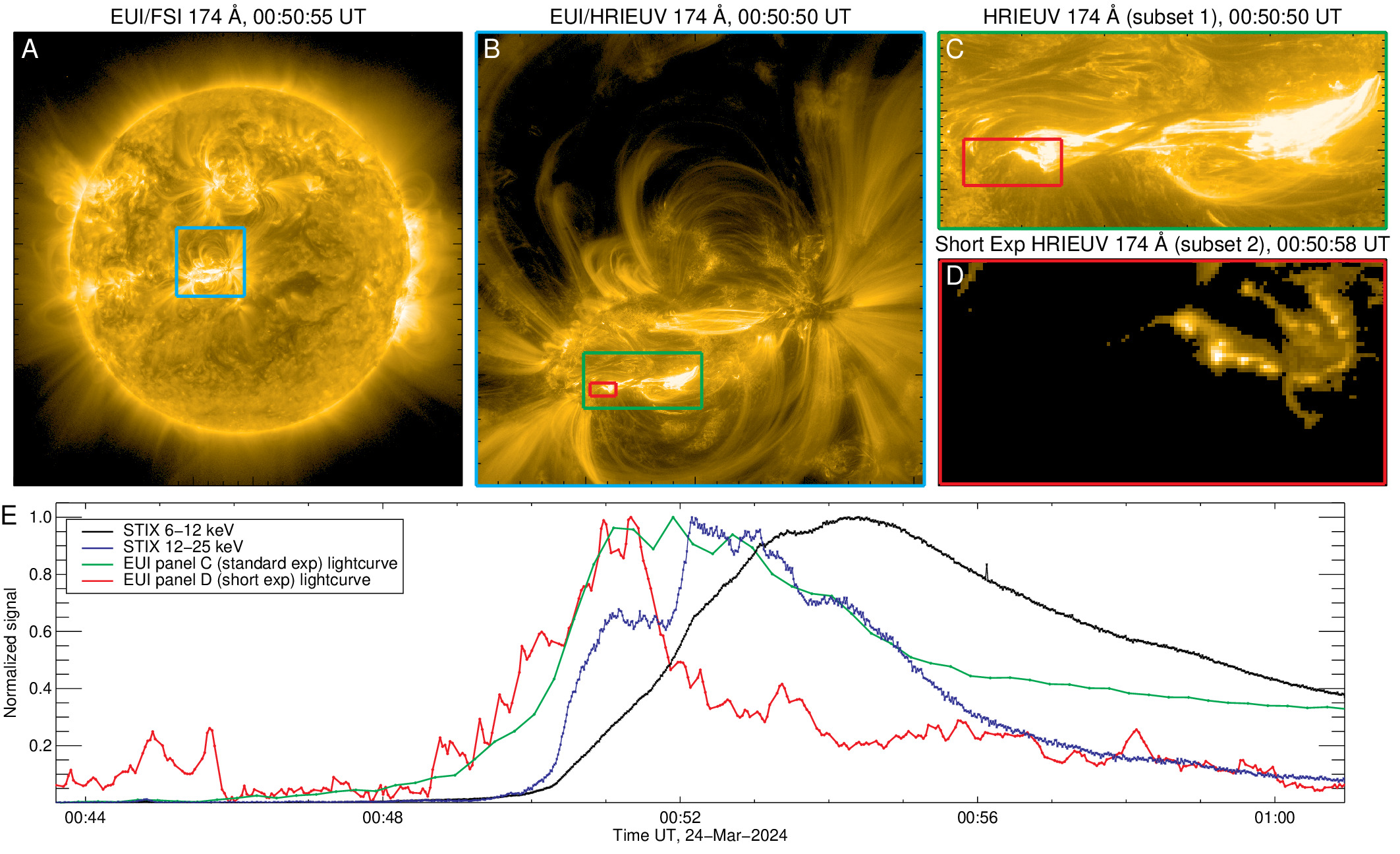}
\caption{
Overview of the 2024 March 24th C9.9-class solar flare. A) Full Sun EUI/FSI 174 \AA\ image, cropped on the inner corona. The cyan FOV marks the EUI/HRIEUV FOV. B) Full (standard exposure) EUI/HRIEUV image, capturing the solar flare within active region AR 13615 (marked by the green box). C) Cropped (standard exposure) EUI/HRIEUV image, within the green FOV marked in panel B. D) Cropped short-exposure EUI/HRIEUV image, capturing the flare ribbons within the red FOV in panels B and C. E) X-ray and EUV time series of the solar flare, including STIX 6-12 and 12-25 keV emission, and light curves of (standard exposure) EUI/HRIEUV 174 \AA\ emission within the FOV of panels C (standard-exposure) and D (short exposure). Panel E lightcurves are normalized between minimum and maximum values.
}
\label{fig:overview}
\end{figure*}

The solar flare was observed by EUI and Spectrometer Telescope for Imaging X-rays \citep[STIX,][]{Krucker2020}. Figure \ref{fig:overview}A presents an 174 \AA\ extreme ultraviolet (EUV) image of the Sun from the EUI Full Sun Imager (FSI) at the time of the flare, cropped to the inner solar corona. The blue box outlines the field-of-view (FOV) of the EUI High Resolution Imager in the EUV (HRIEUV), a synchronous 174 \AA\ image from which is shown in Figure \ref{fig:overview}B. HRIEUV observes with a 2kx2k pixel are with a pixel scale of 0.492\arcsec, corresponding to 140 km at a distance of 0.38 au. During this event, HRIEUV observed with a sequence of one standard-exposure (2\,s) image, followed by 6 short-exposure (0.04 s) images, at a cadence of 2  s per image. This means that the standard-exposure images are captured at a cadence of 16 s, and the short-exposure images a cadence of 2 seconds, followed by a 4 second gap after every 12 seconds. Short-exposure HRIEUV observations offer a major advantage over most EUV imagers, by avoiding saturation in the brightest flare pixels. An introduction to these observations is provided in \citet{Collier2024}.

For comparison, the Solar Dynamics Observatory's Atmospheric Imaging Assembly \citep[SDO/AIA,][]{Lemen2012} has a pixel size of 435 km, EUV channel cadence of 12 seconds, and experiences major issues with saturation during solar flares. The active region and solar flare were also co-observed by Earth-based instruments \citep[outlined in][]{Ryan2025}, but we do not analyze these data in this work.

The green and red boxes in Figure \ref{fig:overview}B mark the FOV of panels C and D, respectively. Figure \ref{fig:overview}C presents a cropped (standard exposure) HRIEUV snapshot of the impulsive phase of the C9.9-class solar flare. The 174 \AA\ filter contains contributions primarily from \ion{Fe}{9} and \ion{Fe}{10}, sensitive to temperatures in the near 1 MK range.
Bright, saturated flare loops are visible in the image, from the coronal \ion{Fe}{9} and/or \ion{Fe}{10} emission. At the left edge of the flare (marked with the red rectangle), a hook-like flare ribbon structure is visible. This flare ribbon structure, visible without higher-altitude coronal emission blocking its view, likely originates from lower in the solar atmosphere -- either in the lower corona or transition region. HRIEUV 174 \AA\ flare ribbons have previously been reported in \citet{Collier2024}. 

Figure \ref{fig:overview}D shows a short-exposure image of the flare ribbon, cropped to the red box in panels C and D. Because of the short exposure, pixels below a certain intensity threshold have no signal, appearing black in the image. The image provides a view of bead-like structures along the flare ribbon, which we analyze in this study.

The bottom panel in Figure \ref{fig:overview} presents the time series of the flare. The green and red curves show the lightcurves of the green (standard exposure) and red (short-exposure) HRIEUV FOVs in panel C/D, and blue and black curves the 6-12 and 12-25 keV hard X-ray (HXR) emission respectively, observed by the Solar Orbiter STIX instrument. The lightcurves are normalized between their minimum and maximum values during the plotted time range.

\section{Flare Ribbon Analysis} \label{sec:analysis}

\begin{figure*}%[h]
\centering
\includegraphics[width=18cm]{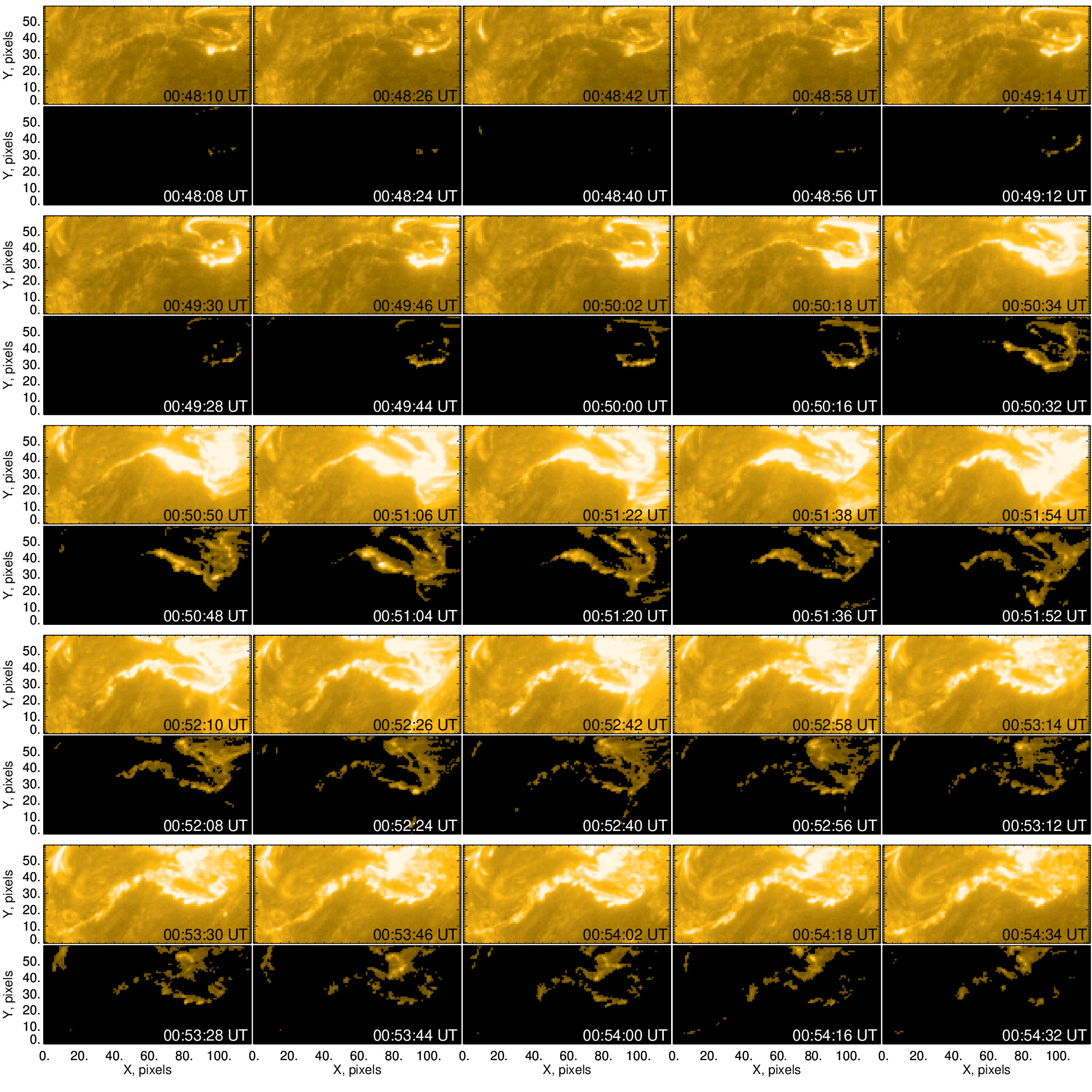}
\caption{
EUI/HRIEUV 174 \AA\ snapshots of solar flare ribbon evolution. Adjacent top/bottom panels show EUI/HRIEUV standard (2 s) and short (0.04 s) exposure images respectively, within the same FOV (the short exposure images precede the standard exposure images by two seconds). Snapshots span from 00:48:10 to 00:54:34 UT. The movie version of the figure shows the same images at full cadence, over the same time range.
}
\label{fig:snapshots}
\end{figure*}

Figure \ref{fig:snapshots} shows the evolution of the flare ribbon within the red FOV outlined in Figure \ref{fig:overview}. Each row of images contains a sequence of pairs of near-contemporaneous images (separated by two seconds) with a standard-exposure (top) and short-exposure (bottom). The ribbon image sequence spans over the time period from 00:48:10--00:54:34, covering the impulsive phase of the flare to the peak of 6-12 keV HXR emission. The image pairs show the standard-exposure images at maximum cadence (16 s), but only 1-in-5 short-exposure images (to match the 16 s cadence of the standard exposure images). The movie version of the figure shows the images at full cadence.

At the start of the flare, small bright beads are visible within the standard-exposure images. By 00:49:30 UT, the flare ribbon has become saturated in the standard-exposure image, but the fine bead-like features remain visible in the short-exposure images. Observing detail within an otherwise-saturated region is a clear advantage of observing flares with short-exposure imagery. The short-exposure beads can be tracked up until at least 00:53:00\,UT, where the lowering intensity creates gaps in signal along the ribbons. Around this same time (as designed by the exposure time ratio between observing modes), the flare ribbon structures fall below the saturation limit of the standard-exposure images, where we can continue to monitor them. From $\approx$00:53\,UT onwards, the beads evolve into more `candle flame' or saw-tooth-like structures, with apparent height variation along the ribbon. The brightest of these are also visible in the short-exposure images. Because of the single viewpoint of the EUI images, it is not possible to fully distinguish these candle-flame features as horizontal or vertical variations. However, because there is little horizontal ribbon motion at this time, it is more likely they are vertical features, rather than horizontal structures forming at the leading or receding edge of the ribbon. By the last panels of the figure, any notable features along the flare ribbon have dissipated.

\subsection{Ribbon Tracking} 

\begin{figure*}%[h]
\centering
\includegraphics[width=9cm]{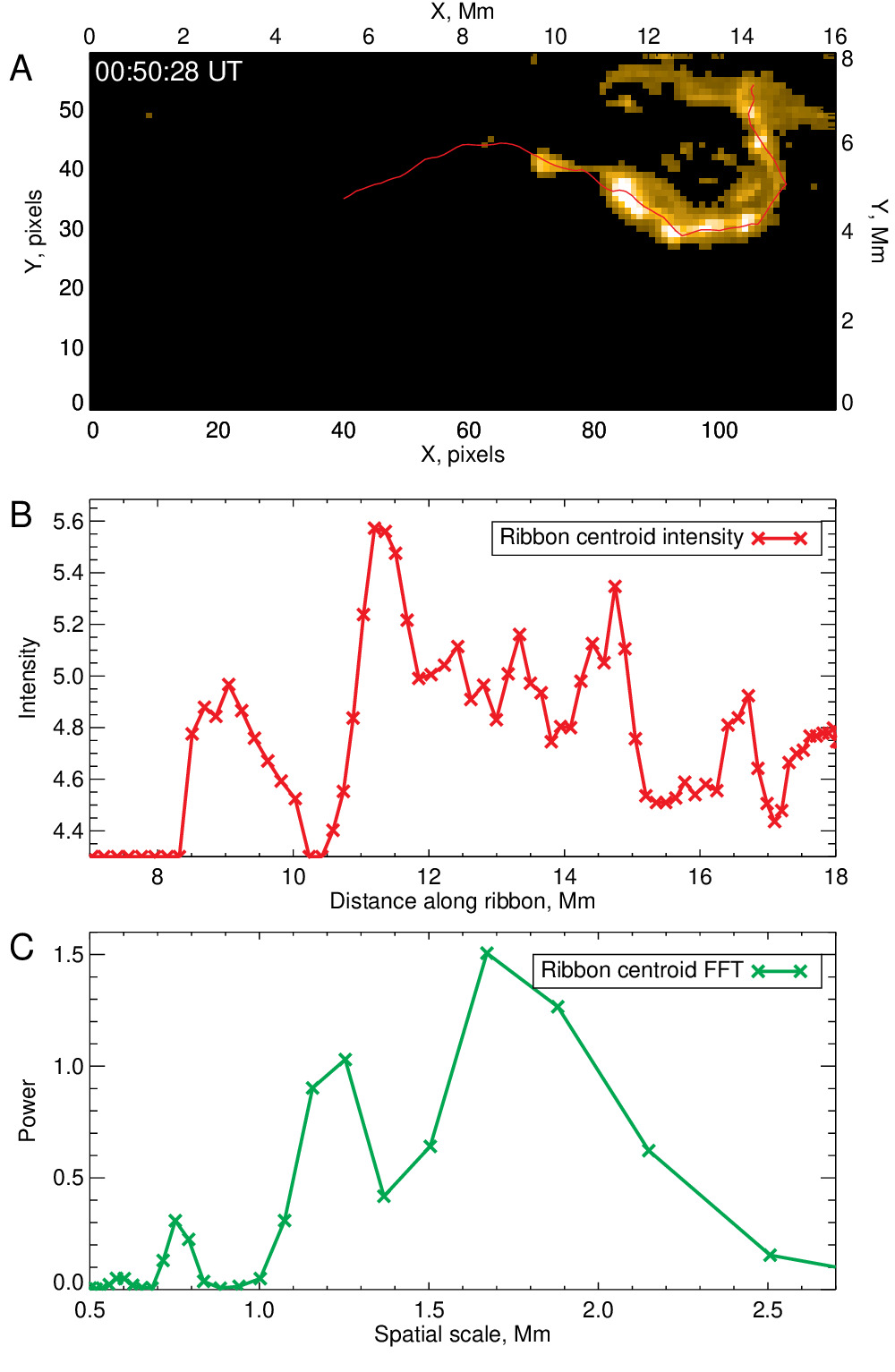}
\caption{
Demonstration of flare ribbon tracking. A) Sample short-exposure EUI/HRIEUV image, with the red line marking the tracking central ribbon axis along the flare ribbon. B) Intensity cross-section along the central ribbon axis shown in panel A. C) Spatial Fast Fourier Transform (FFT) of the intensity cross-section presented in panel B. 
}
\label{fig:tracking}
\end{figure*}

We implement flare ribbon tracking to track the central ribbon axis of the flare ribbon as it evolves in time and space, with an example presented in Figure \ref{fig:tracking}. The tracking routine is applied to the standard-exposure images, semi-automatically, and works as follows:
\begin{itemize}\setlength\itemsep{0em}
    \item \textbf{Manually define mask}: For three images throughout the ribbon evolution, we manually select pixels along the central ribbon axis to define its location. The mask is the set of all pixels within 7 pixels radially from the ribbon location at this time. As the ribbon exits a mask, a new mask is set. Overall, we used three masks, as this was enough to ensure that the evolving ribbon structure was always visible within the sequence of pre-set masks.
    \item \textbf{Positions of maximum intensity}: For each time step, we take cross-sections in each pixel perpendicular to the central mask axis, and record the brightest pixel in each cross-section. This creates a noisy array of pixels close to the brightest central ribbon axis. In the case of multiple saturated pixels in a given cross-section (where the maximum intensities are equal), we select the center of these pixels.
    \item \textbf{Smoothed central ribbon axis}: For each time step, a smoothed moving-average is applied to the noisy array of maximum-intensity pixels, to create a smooth central axis along the flare ribbon. The result is a central ribbon axis that tracks the evolving ribbon structure with time. To retain information on the relative location of features along the flare ribbon, the ribbon central axis length is fixed throughout all images. This means that for earlier images, where the ribbon is short, the majority of the central axis consists of NaN values (at the location of the central mask axis). 
    \item \textbf{Interpolate}: For short-exposure images, the central ribbon axis is interpolated between the central axis locations of adjacent standard-exposure images.
\end{itemize}

Figure \ref{fig:tracking}A shows an example short-exposure image, with the red line marking the location of the tracking central ribbon axis, which successfully represents the ribbon structure (including the bright ribbon beads). A manual inspection is undertaken across the full ribbon evolution, to confirm that the central ribbon axis adequately represents the position and shape of the flare ribbon.

With a successful routine to track the evolution of the central ribbon axis with time, we can measure intensity variations along the flare ribbon for any image. For each pixel along the central ribbon axis, we take the intensity as the mean of a 5x5 array centered around that pixel. We do this to alleviate the impact of errors arising from individual pixel-to-pixel variations in the ribbon tracking, and better represent the ribbon intensity beyond that found in a single pixel at the ribbon center. Figure \ref{fig:tracking}B shows the intensity cross-section along the image tracking example in Figure \ref{fig:tracking}A, revealing spikes in intensity along the ribbon. These intensity spikes match the locations of the bead-like structures visible in the image. Many of the adjacent beads have (peak-to-peak) separations of 3-6 pixels, corresponding to bead separation distances of $\approx420-840$km (1 pixel = 140 km).

\subsection{Ribbon Sub-Structure}

\begin{figure*}%[h]
\centering
\includegraphics[width=18cm]{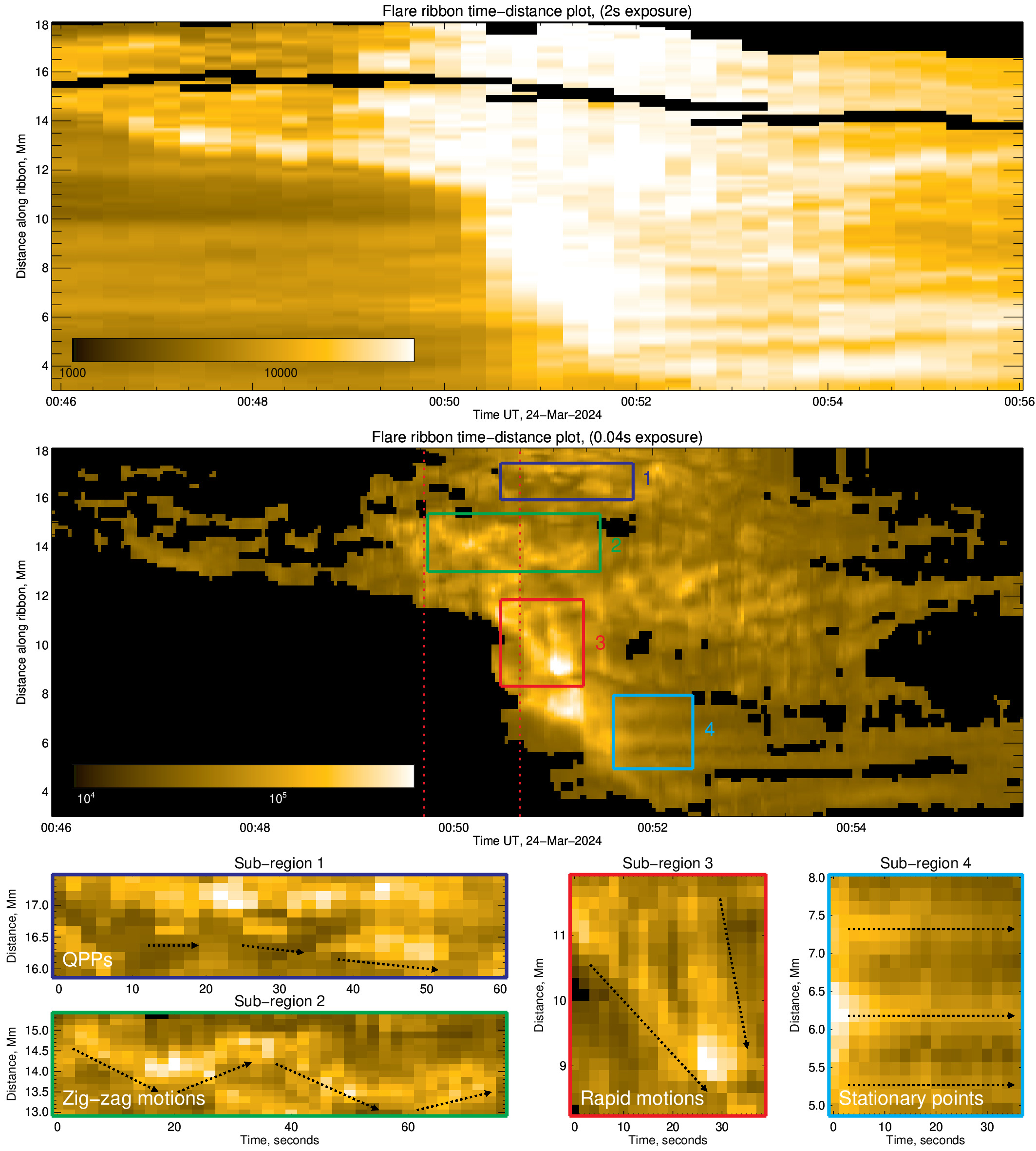}
\caption{
Time-distance plot of intensity variations along the central ribbon axis, from standard exposure (top panel) and short exposure (middle panel) HRIEUV images. The dotted red vertical line in the middle panel marks the earliest start times and end time of exponential growth, as presented in Figure \ref{fig:FFT}.
The bottom panels present cropped sub-regions 1-4 from the middle panel, highlighting 1) QPP-like brightenings, 2) `zig-zag' motions, 3) rapid motions, and 4) stationary bright points. 
}
\label{fig:stack}
\end{figure*} 

By combining the intensity cross-section measured for every time step, we create the time-distance stack plots in Figure \ref{fig:stack}. The top and middle panels of Figure \ref{fig:stack} show the ribbon intensity stack plot from the standard and short-exposure HRIEUV images, respectively. The standard exposure stack plot shows the growth of the ribbon, before the region becomes saturated. From 00:52\,UT onwards, the saturation begins to subside, providing a view of the spatially-distinct bead-like structures along the ribbon (most prominent between 4-9 Mm). Although the short-exposure stack plot lacks the signal to capture the early growth of the ribbon, the faster cadence and lack of saturation provides a much better view of evolving sub-structure within the flare ribbon. The stack plot reveals a mixture of multiple bright points moving in different patterns and speeds. We highlight four behaviors in particular, labeled sub-regions 1-4 in Figure \ref{fig:stack}. The sub-regions are highlighted by colored boxes in the short-exposure stack plot, and presented in a zoomed-in view in the bottom four panels of the figure. The distinct sub-region behaviors are:

\begin{itemize}
    \item \textbf{Sub-region 1}: Beads exhibit quasi-periodic `on-off' intensity pulsations as they travel along the flare ribbon. The pulsations continue beyond the sub-region FOV.
    
    \item \textbf{Sub-region 2}: Beads exhibit back-and-forth `zig-zag' motion as they flow along the flare ribbon. This manifests itself as a `W' shape within the stack plot, extending beyond the sub-region FOV. 
    
    \item \textbf{Sub-region 3}: Rapid bead motions along the ribbon, at speeds of $\approx 600$ km/s. Two beads coalesce to form a brighter one.
    
    \item \textbf{Sub-region 4}: Stationary bright points at a fixed position in the ribbon, persisting for 4+ minutes.
\end{itemize}

\subsection{Spatial Fourier Analysis of Flare Ribbons} 

We search for systematic spatial frequencies along the ribbon with FFT analysis of the intensity cross-section. FFTs are commonly applied to time series, but here we apply it to a spatial cross-section at a fixed time, to search for the spatial scales present at a given time step. Before applying the FFT, we detrend the timeseries (over 60 pixels) using a 2nd-degree polynomial Savitzky–Golay filter \citep{savgol_1964}, and apply a Hanning window. Figure \ref{fig:tracking}C shows an example FFT for the cross-section presented in Figure \ref{fig:tracking}B. The FFT reveals strong FFT power peaks at spatial scales of 1.25 and 1.67 Mm. Smaller peaks are also visible at $\approx$570 and 750 km, close to the closest separations that we manually identified within the intensity cross-section.

\begin{figure*}%[h]
\centering
\includegraphics[width=18cm]{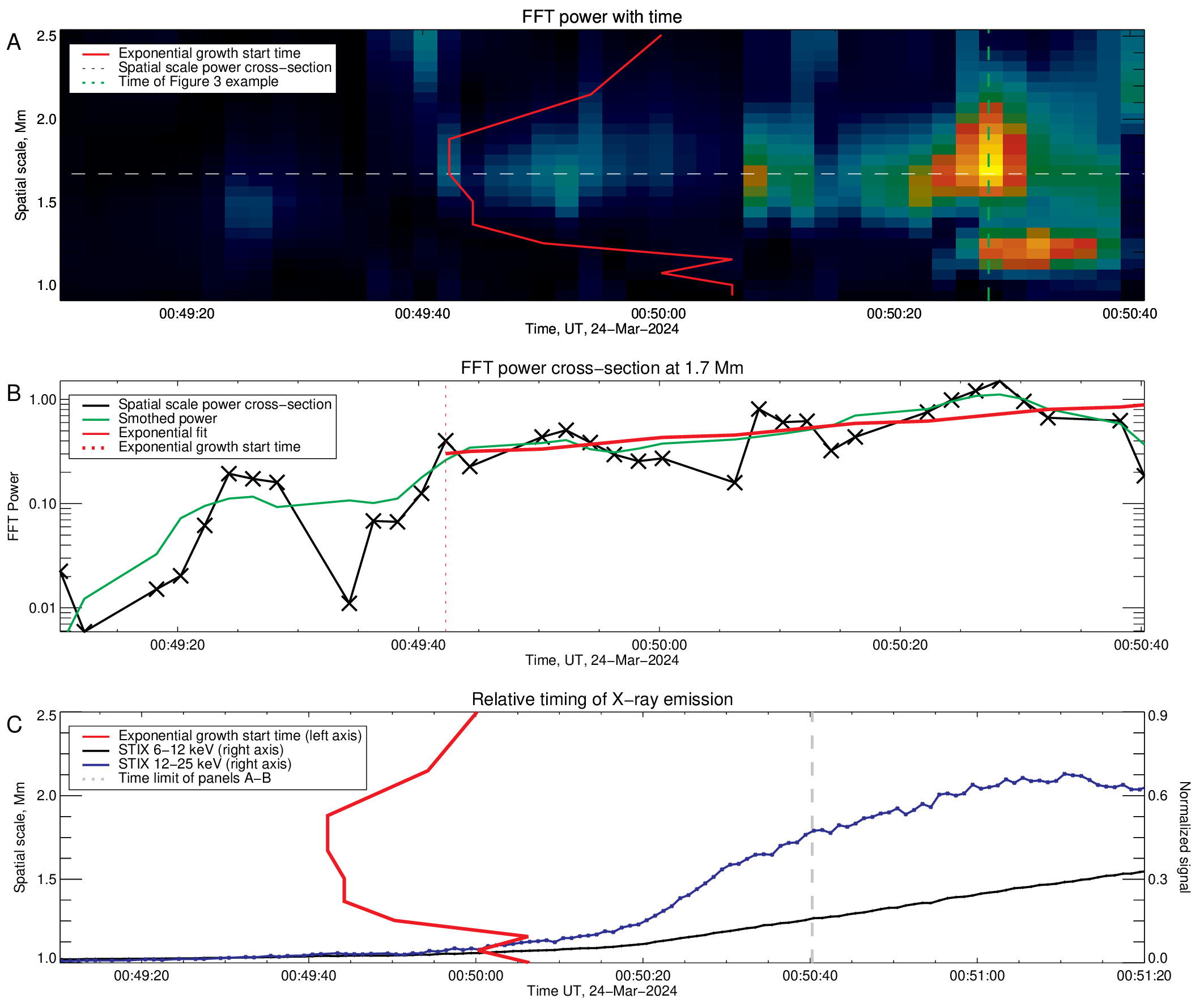}
\caption{
Spatial FFT Analysis. A) Time-FFT plot, showing FFT power along the flare ribbon at each time step. The red line marks the start time of exponential growth at each spatial scale. The horizontal gray dashed line shows the spatial scale power cross-section presented in panel B. The vertical green line marks the time of the example ribbon image used to demonstrate the tracking and FFT procedure in Figure \ref{fig:tracking}.
B) Cross-section of the panel A Time-FFT plot, at a spatial scale of 1.7 Mm. The green line shows the smoothed growth curve, and solid red line the best exponential fit to the smoothed data (with a growth rate of $\approx$ 0.01 s$^{-1}$). The vertical dashed red line marks the time exponential growth starts at this spatial scale.
C) The relative timing of exponential growth at different spatial scales (red curve, left axis), relative to the onset of HXR emission (blue/black curves, right axis). The vertical dashed gray line marks the end time of panels A and B.
}
\label{fig:FFT}
\end{figure*} 

We calculate the FFT spectrum of the  central ribbon axis intensity at each time step. Following this, we combine them to create a stack plot of FFT power with time. A cropped version of this is presented in Figure \ref{fig:FFT}A, with a sub-range of the flare duration on the x-axis (cropped to the main growth phase of the ribbon), with spatial scale along the y-axis. The color scale denotes the FFT power. The vertical dashed green line on Figure \ref{fig:FFT}A marks the time of the example ribbon frame used for the tracking and FFT demonstration in Figure \ref{fig:tracking}. The FFT peaks of Figure \ref{fig:tracking}C at 1.25 and 1.67 Mm are visible in the power stack plot at this time.

Taking a horizontal slice along the FFT stack plot in Figure \ref{fig:FFT}A provides the growth of power with time at a given spatial scale. Figure \ref{fig:FFT}B shows an example power growth plot for a scale of 1.7 Mm, marked by the dashed line in Figure \ref{fig:FFT}A. The black curve marks the power growth, and green curve shows a smoothed moving average. To measure the period of exponential growth within the power time series, we fit a linear curve to the logarithm of the smoothed power data. We fix the end point of exponential growth, and fit sequential linear curves to different portions of the time series, ranging from different start times to the assigned end time. Out of this selection of linear fits, we pick the fit with the start time that yielded the lowest least-squares residual. For this fit, we record its start time as the start of exponential growth at that specific spatial scale. By repeating this process for every spatial scale across the power spectrum, we produce an array of the varying start times of exponential growth at each spatial scale. This array is plotted as the red curve in Figure \ref{fig:FFT}A. By taking the ratio of the exponential power growth to exponential growth duration, we can also calculate the growth rate at each spatial scale.

We find the exponential growth begins at a key spatial scale of around 1.7 Mm, before the subsequent growth of progressively larger and smaller space scales. Figure \ref{fig:FFT}C compares the timing of the spatial scale growth with the onset of 6-12 and 12-25 keV HXRs.
We find growth rates in the range of 0.01-0.06 s$^{-1}$, with the lowest growth rate occurring at the spatial scale of earliest growth onset.
% We find that the growth precedes notable 12-25 keV HXR onset by 10-15 s. 

\section{Discussion} \label{sec:disc}

The fast-cadence and non-saturated short-exposure EUI/HRIEUV observations reveal a mixture of different behaviors along the flare ribbon. These behaviors all occur over distinct timescales, with different sizes, apparent velocities and durations. This suggests that the range of flare ribbon substructure behavior observed in this flare (and in other studies) can most likely \textit{not} be attributed to a single process in the flare current sheet, but are instead caused by the complex myriad of overlapping and intersecting processes. 
It is important to note that the velocities observed in EUV flare ribbon studies are not representative of physical flows along the flare ribbons, but rather apparent motions resulting from changes in connectivity within the flaring corona, where the bright ribbon regions mark recent locations of magnetic connectivity.

The stationary bright points observed in sub-region 4 of Figure \ref{fig:stack} bear much resemblance to the lower altitude \textit{blobs} recently reported in ground-based observations \citep[e.g.][]{Thoen2025,Yaduv2025}, occurring on comparable spatial scales ($\approx 1.2$ Mm) and persisting (with limited POS motion) for durations of several minutes. Spectropolarimetric \ion{Ca}{2} measurements from \citet{Yaduv2025} find evidence of localized heating in the blobs, consistent with MHD predictions \citep[e.g. from][]{Dahlin2025} of current sheet tearing during
flare-associated magnetic reconnection. They note, however, that their observations lack the temporal resolution needed to study the behavior of these features in more detail. In our work, we can utilize high resolution images (albeit without any spectroscopic or spectropolarimetric data) to further analyze ribbon fine structure. 

In contrast to the slow moving blobs are the much faster flare kernels shown in the sub-region 3 (Figure \ref{fig:stack}). These features, with apparent motions around 600 km/s, are similar in appearance to the 450 km/s running motions detected by \citet{Lorincik2019} (and references therein), which are common ribbon signatures interpreted as evidence of sub-Alfv\'enic slipping reconnection \citep[see][]{Priest1992,Aulanier2006}. Evidence for super-Alfv\'enic slipping reconnection has also been presented in \citet{Lorincik2025}, with apparent ribbon motions detected to be as high as $2652\pm1041$km. We do not detect clear evidence for apparent velocities this fast, despite having sufficient cadence to detect them.

Also highlighted in Figure \ref{fig:stack} are the dynamic behaviors shown in sub-regions 1 and 2. Sub-region 1 shows quasi-periodic brightenings of the bead structures as they propagate along the ribbon, with periods around $\sim15-30$ seconds. Quasi-periodic pulsations (QPPs) are oscillations in a solar flare time series, typically with periods ranging from tens of seconds to tens of minutes. There are several candidate processes potentially responsible for generating QPPs, including bursty magnetic reconnection (such as that created in a tearing plasmoid-mediated current sheet), various MHD modes, or a combination of the two \citep[see reviews by][]{nakariakov2009,vandoorsselare2016,McLaughlin2018,Kupriyanova2020,Zimovets2021}.
QPPs have been detected across the full electromagnetic spectrum, in both the flaring corona and chromospheric flare ribbons. The small-scale periodic ribbon brightenings presented in this work are QPP-like in nature, occurring within a single bright ribbon kernel over a period of $\sim15-30$ seconds. Previous analysis of oscillations within flare ribbon properties \citep[e.g.][]{Brannon2015,Brosius2015,Corchado2024,French2025b} have attributed the QPP origins to waves or turbulence in the coronal current sheet, consistent with that expected of the tearing-mode instability or bursty magnetic reconnection. This occurs as periodic reconnection at the top of the flare arcade leads to periodic energy deposition in the chromosphere, resulting in the varying intensity of the ribbon bead structure. We do note, however, that many bright beads along the ribbon do not pulsate in this way, and these clear intensity oscillations are only seen in a small number of bright points. 

Sub-region 2 of Figure \ref{fig:stack} displays a different kind of oscillatory behavior. Instead of revealing oscillations in bead intensity with time, we see an oscillation in the position of the bright point. Whilst flowing along the ribbon, the bead exhibits quasi-periodic `back-and-forth' motions; briefly changing direction before continuing in its original direction. This manifests itself as a `W' profile in Figure \ref{fig:stack}. This behavior is unusual, and to our best knowledge has not been reported before. The origin of this behavior is challenging to describe qualitatively, but one potential explanation falls to modeling efforts from \citet{Wyper2021}, which simulates observable flare ribbon structures originating from the tearing-mode instability in the coronal current sheet. Figure 9 of \citet{Wyper2021} shows a simulation of fine-scale hook-like structures in the flare ribbon current maps. As they propagate with time, the hook-like structures curl up like a breaking ocean wave, before returning to a simpler structure. In observations, such small-scale hooks have not been found along the primary flare ribbon body. However, it is possible that these hook structures may manifest themselves as bead structures in observations. This could either be due to the limited resolution of solar telescopes failing to resolve the fine structure, or due to local thermodynamic conditions `blending' the hooks into the smoother beads. It is also possible that the shape of the structures could vary with height in the solar atmosphere. A qualitative interpretation of the scenario demonstrated by \citet{Wyper2021} would suggest that for structures traveling along the ribbon, the `curling' or `wave breaking' pattern of the hooks would result in a slight backwards movement opposite to the direction of movement -- similar to the behavior of the beads in our observations. In \citet{Wyper2021}, this behavior is attributed to the tearing-mode instability within the flare current sheet, and thus a possible further indication for the presence of the process in the event presented in this work. 

Figure \ref{fig:FFT} conducts a more quantitative search for evidence of the tearing mode instability. Utilizing methodology from \citet{French2021} \citep[inspired by magnetospheric studies of][]{Kalmoni2015,Kalmoni2018}, we track the start time of exponential growth at different spatial scales along the central ribbon axis. 
During the tearing mode instability in simulation work, the current sheet tears at an initial spatial scale, forming magnetic island plasmoid structures along the current sheet. As magnetic reconnection progresses, plasmoids continue to tear apart into smaller structures, whilst magnetic reconnection at boundaries of adjacent magnetic structures allows the simultaneous coalescence into larger spatial scales. This results in a simultaneous cascade and inverse cascade of spatial scales in the current sheet \citep{Tenerani2020}. Similar to the results of \citet{French2021}, we find evidence for this process within the flare ribbons presented in this work. Figure \ref{fig:FFT} demonstrates flare ribbon structure to originate at a scale of 1.7-1.9 Mm at the beginning of flare onset (prior to notable HXR emission), before the subsequent growth of progressively larger and smaller scales. \citet{French2021} analyze two flare ribbons independently to prove that the change in spatial scales are originating from the coronal current sheet connecting the two, but we lack observations of the second flare ribbon to complete the same verification in this flare. Nonetheless, repeating the discovery of this pattern of spatial scale growth provides firm quantitative evidence for the presence of the tearing mode instability. Curiously, the spatial scales at which the flare ribbon structures first manifests themselves at during flare onset are remarkably similar between this work and \citet{French2021} ($\approx 1.75$ Mm), despite a stark difference in flare magnitude (C9.9-class versus B-class), observed by different instrumentation (EUI/HRIEUV versus IRIS/SJI). This flare ribbon scale is dependent on the scale of current sheet tearing in the corona, but also determined by the coronal energy release volumes, levels of magnetic flux, and differences in flux tube areas between the corona and chromosphere. We therefore believe that the repeated detection of the $\approx$1.75 Mm value is likely a coincidence, and not fundamental to the current sheet tearing processes. The exponential growth rates measured in this study are in the range of $0.01-0.06$ s$^{-1}$, a factor of $\approx4$ lower than those detected in the smaller flare analyzed in \citet[]{French2021}.

\section{Conclusions} \label{sec:concl}

We have presented novel observations from the Solar Orbiter EUI/HRIEUV instrument at 174 \AA. We analyze fast cadence (2s) flare ribbon observations, at a spatial resolution three times finer than AIA (and comparable to the temporal and spatial resolution of IRIS SJI observations). Crucially, due to the short-exposure mode of the instrument, the data is unsaturated, allowing the analysis of fine bead-like structures within the flare ribbon. Utilizing flare ribbon tracking, we detect a variety of simultaneous flare ribbon behaviors, most likely originating from multiple processes in the coronal current sheet. Many of these ribbon behaviors have been observed independently before, but never before together all at once.
Although we do not believe a single process can account for all the ribbon behavior, many observed ribbon behaviors are consistent with the presence of the tearing mode instability. These include QPP-like brightenings, persistent blob-like features, and previously-unreported zig-zag motions of bright bead-like kernels along the central ribbon axis. Finally, a quantitative FFT analysis of the spatial separation of beads along the flare ribbons reveals spatial scales beginning at 1.7-1.9 Mm, before the subsequent growth of progressively larger and smaller spatial scales. This discovery, also validating earlier works, is a key prediction of the tearing mode instability.

\begin{acknowledgments}
R.J.F. thanks support from NASA HGI award 80NSSC25K7927. R.J.F. also thanks support from the Royal Observatory of Belgium Guest Investigator Program. M.D.K. and R.Y. acknowledge support by NASA ECIP award 80NSSC19K0910 and NSF CAREER award SPVKK1RC2MZ3. D.B., E.D, M.D. and D.C.T. acknowledge support fro BELSPO in the framework of the ESA-PRODEX program, PEA SIDEX 4000145189. R.P. is supported by NASA Solar Orbiter Guest Investigator Grant Number: 80NSSC24K1243.  C.T. is supported by NSF grant number 2407850.
\end{acknowledgments}

\bibliography{bibliography}{}
\bibliographystyle{aasjournalv7}



\end{document}